\documentclass[11pt,a4paper]{article}
\usepackage[hyperref]{acl2021}
\usepackage{times}
\usepackage{latexsym}
\usepackage{booktabs}
\usepackage{mathtools}
\usepackage{todonotes}

\usepackage{amsmath}
\usepackage{amssymb}
\usepackage{ifsym}
\DeclareMathOperator*{\argmax}{argmax}
\DeclareMathOperator*{\getentity}{\mathtt{entity}}

\newcommand{\moleman}{{\sc moleman}}
\newcommand{\modelf}{Model F}
\newcommand{\modelfplus}{Model F$^+$}
\newcommand{\mgenre}{{\sc mGENRE}}

\usepackage{microtype}

\aclfinalcopy 
\def\aclpaperid{2604} 

\let\norm\undefined 
\DeclarePairedDelimiter\norm{\lVert}{\rVert}

\title{MOLEMAN: Mention-Only Linking of Entities with a Mention Annotation Network}

\author{Nicholas FitzGerald, Jan A. Botha, Daniel Gillick, Daniel M. Bikel, \\ \textbf{Tom Kwiatkowski, Andrew McCallum}\\
  Google Research \\
  \texttt{\{nfitz,jabot,dgillick,dbikel,tomkwiat,mccallum\}@google.com}
 }

\date{}

\begin{document}
\maketitle
\begin{abstract}

We present an instance-based nearest neighbor approach to entity linking. In contrast to most prior entity retrieval systems which represent each entity with a single vector, we build a contextualized mention-encoder that learns to place similar \emph{mentions} of the same entity closer in vector space than mentions of different entities. This approach allows all mentions of an entity to serve as ``class prototypes'' as inference involves retrieving from the full set of labeled entity mentions in the training set and applying the nearest mention neighbor's entity label. Our model is trained on a large multilingual corpus of mention pairs derived from Wikipedia hyperlinks, and performs nearest neighbor inference on an index of 700 million mentions. It is simpler to train, gives more interpretable predictions, and outperforms all other systems on two multilingual entity linking benchmarks.

\end{abstract}

\section{Introduction}

A contemporary approach to entity linking represents each entity with a textual description $d_e$, encodes these descriptions and contextualized mentions of entities, $m$, into a shared vector space using dual-encoders $f(m)$ and $g(d_e)$, and scores each mention-entity pair as the inner-product between their encodings~\citep{botha2020entity,wu2019scalable}.
By restricting the interaction between $e$ and $m$ to an inner-product, this approach permits the pre-computation of all $g(d_e)$ and fast retrieval of top scoring entities using maximum inner-product search (MIPS).

\begin{figure*}[h]
    \centering
    \includegraphics[width=\textwidth]{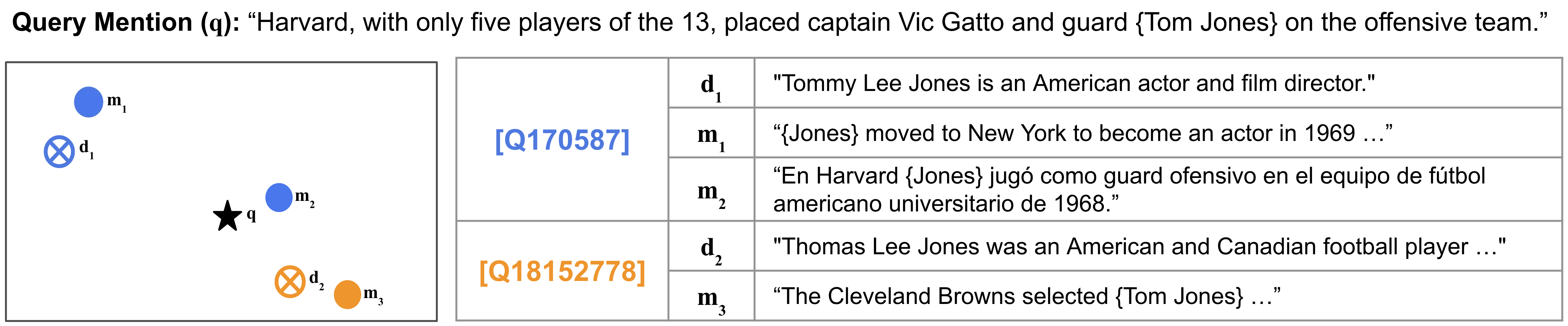}
    \caption{\small Illustration of hypothetical contextualized mention ($m$) and multilingual description ($d$) embeddings for the entities `Tommy Lee Jones (Q170587)' and `Tom Jones (Q18152778). The query mention [$\bigstar$] pertains to the former's college football career, which is unlikely to be captured by the high-level entity description. A retrieval against descriptions would get this query incorrect, but with indexed mentions gets it correct. Note that prior dual-encoder models that use a single vector to represent each entity are forced to contort the embedding space to solve this problem.}
    \label{fig:teaser}
\end{figure*}

Here we begin with the observation that many entities appear in diverse contexts, which may not be easily captured in a single high-level description.
For example, Actor Tommy Lee Jones played football in college, but this fact is not captured in the entity description derived from his Wikipedia page (see Figure~\ref{fig:teaser}). 
Furthermore, when new entities need to be added to the index in a zero-shot setting, it may be difficult to obtain a high quality description.
We propose that both problems can be solved by allowing the entity mentions themselves to serve as exemplars.
In addition, retrieving from the set of mentions can result in more interpretable predictions -- since we are directly comparing two mentions -- and allows us to leverage massively multilingual training data more easily, without forcing choices about which language(s) to use for the entity descriptions.

We present a new approach (\moleman{}\footnote{Mention Only Linking of Entities with a Mention Annotation Network}) that maintains the dual-encoder architecture, but with the same mention-encoder on both sides.
Entity linking is modeled entirely as a mapping between mentions, where inference involves a nearest neighbor search against all known mentions of all entities in the training set.
We build \moleman{} using exactly the same mention-encoder architecture and training data as \modelf{} \cite{botha2020entity}. 
We show that \moleman{} significantly outperforms \modelf{} on both the Mewsli-9 and \newcite{tsai-roth-2016-cross} datasets, particularly for low-coverage languages, and rarer entities. 

We also observe that \moleman{} achieves high accuracy with just a few mentions for each entity, suggesting that new entities can be added or existing entities can be modified simply by labeling a small number of new mentions.
We expect this update mechanism to be significantly more flexible than writing or editing entity descriptions.
Finally, we compare the massively multilingual \moleman{} model to a much more expensive English-only dual-encoder architecture \cite{wu2019scalable} on the well-studied TACKBP-2010 dataset~\citep{ji2010overview} and show that \moleman{} is competitive even in this setting.

\section{Overview}

\paragraph{Task definition}

We train a model that performs entity linking by ranking a set of entity-linked \emph{indexed mentions-in-context}.
Formally, let a mention-in-context $\mathbf{x} = [x_1, ..., x_n]$ be a sequence of $n$ tokens from vocabulary $\mathcal{V}$, which includes designated entity span tokens. 
An \emph{entity-linked} mention-in-context $m^i = (\mathbf{x}^i, e^i)$ pairs a mention with an entity from a predetermined set of entities $\mathcal{E}$.
Let $\mathcal{M_I} = [m^1, ..., m^k]$ be a set of entity-linked mentions-in-context, and
let $\getentity(\cdot) : \mathcal{M_I} \rightarrow \mathcal{E}$ be a
function that returns the entity $e^i \in \mathcal{E}$ associated with $m^i$, and $\mathtt{x}(\cdot)$ returns the token sequence $\mathbf{x}^i$.

Our goal is to learn a function $\phi(m)$ that maps an arbitrary mention-in-context token sequence $m$ to a fixed vector $\mathbf{h}_m \in \mathcal{R}^d$ with the property that 
\begin{equation*}
    y^* = \getentity\left(\argmax_{m' \in \mathcal{M}_I}[\phi(\mathtt{x}(m'))^T \phi(\mathbf{x}_q)]\right)
\end{equation*}
gives a good prediction $y^*$ of the true entity label of a query mention-in-context $\mathbf{x}_q$.

\section{Method}

\subsection{Model}

Recent state-of-the-art entity linking systems employ a dual encoder architecture, embedding mentions-in-context and entity representations in the same space.
We also employ a dual encoder architecture but we score mentions-in-context (hereafter, mentions) against other mentions, with no consolidated entity representations.
The dual encoder maps a pair of mentions $(m, m')$ to a score:
\begin{equation*}
    s(m, m') = \frac{\phi(m)^T \phi(m')}{\norm{\phi(m)}\norm{\phi(m')}}
\end{equation*}
where $\phi$ is a learned neural network that encodes the input mention as a $d$-dimensional vector.

As in~\citep{fevry2020empirical} and~\citep{botha2020entity}, our mention encoder is a 4-layer BERT-based Transformer network~\citep{vaswani2017attention,devlin2019bert} with output dimension $d=300$.

\subsection{Training Process}
\subsubsection{Mention Pairs Dataset}

We build a dataset of mention pairs using the 104-language collection of Wikipedia mentions as constructed by~\newcite{botha2020entity}. 
This dataset maps Wikipedia hyperlinks to WikiData~\citep{vrandevcic2014wikidata}, a language-agnostic knowledge base. We create mention pairs from the set of all mentions that link to a given entity.

We use the same division of Wikipedia pages into train and test splits used by~\newcite{botha2020entity} for compatibility to the TR2016 test set~\citep{tsai-roth-2016-cross}.
We take up to the first 100k mention pairs from a randomly ordered list of all pairs regardless of language, yielding 557M and 31M training and evaluation pairs, respectively.
Of these, 69.7\% of pairs involve two mentions from different languages.
Our index set contains 651M mentions, covering 11.6M entities.

\subsubsection{Hard Negative Mining and Positive Resampling}
\label{sec:hard-negatives}

Previous work using a dual encoder trained with in-batch sampled softmax has improved performance with subsequent training rounds using an auxiliary cross-entropy loss against hard negatives sampled from the current model~\citep{gillick2019learning,wu2019scalable,botha2020entity}.
We investigate the effect of such negative mining for \moleman{}, controlling the ratio of positives to negatives on a per-entity basis.
This is achieved by limiting each entity to appear as a negative example at most 10 times as often as it does in positive examples, as done by~\newcite{botha2020entity}.

In addition, since \moleman{} is intended to retrieve the \emph{most similar} indexed mention of the correct entity, we experiment with using this retrieval step to resample the positive pairs used to construct our mention-pair dataset for the in-batch sampled softmax, pairing each mention $m$ with the highest-scoring other mention $m'$ of the same entity in the index set.
This is similar to the index refreshing that is employed in other retrieval-based methods trained with in-batch softmax~\citep{guu2020realm,lewis2020pre}.

\subsubsection{Input Representations}
Following prior work~\citep{wu2019scalable,botha2020entity}, our mention representation consists of the page title and a window around the mention, with special mention boundary tokens marking the mention span.
We use a total context size of 64 tokens.

Though our focus is on entity mentions, the entity descriptions can still be a useful additional source of data, and allow for zero-shot entity linking (when no mentions of an entity exist in our training set).
We therefore experiment with adding the available entity descriptions as additional ``pseudo-mentions''.
These are constructed in a similar way to the mention representations, except without mention boundaries.
Organic and psuedo-mentions are fed into BERT using distinct sets of token type identifiers.
We supplement our training set with additional mention pairs formed from each entity's description and a random mention, adding 38M training pairs, and add these descriptions to the index, expanding the entity set to 20M.

\subsection{Inference}

For inference, we perform a distributed brute-force maximum inner product search over the index of training mentions.
During this search, we can either return only the top-scoring mention for each entity, which improves entity-based recall, or else all mentions, which allows us to experiment with k-Nearest Neighbors inference (see Section~\ref{sec:results-mewsli}).

\section{Experiments}

\subsection{Mewsli-9}
\label{sec:results-mewsli}

Table~\ref{tab:mewsli-overview} shows our results on the Mewsli-9 dataset compared to the models described by~\newcite{botha2020entity}.
\modelf{} is a dual encoder which scores entity mentions against entity descriptions, while \modelfplus{} adds two additional rounds of training with hard negative mining and an auxiliary cross-lingual objective.
Despite using an identically-sized transformer, and trained on the same data, \moleman{} outperforms \modelfplus{} when training only on mention pairs, and sees minimal improvement from a further round of training with hard negative and resampled positives (as described in Section~\ref{sec:hard-negatives}).
This suggests that training \moleman{} is a simpler learning problem compared to previous models which must capture all an entity's diverse contexts with a single description embedding.
Additionally, we examine a further benefit of indexing multiple mentions per entity: the ability to do top-K inference, and find that top-1 accuracy improves by half a point with k=5.

We also compare to the recent \mgenre{} system of~\newcite{decao2021multilingual}, which performs entity linking using constrained generation of entity names.
It should be noted that this work uses an expanded training set that results in fewer zero- and few-shot entities (see~\newcite{decao2021multilingual} Table 3).

\begin{table}[]
\small
\centering
\begin{tabular}{|l|c|c|c|c|c|}
\hline
         & I & HN & R@1                      & R@10 & R@100 \\
         \hline
\;\modelf{}                   & D & N & 63.0  & 91.7  & 97.4  \\
\;\modelfplus{}               & D & Y & 89.4  & 96.4  & 98.2  \\
\;\mgenre{}                   & -- & -- & 90.6  & --     & --     \\
\hline
\textbf{\moleman}               & M & N & 89.5  & 97.4  & 98.3  \\
               & B & N & 89.6  & 98.0  & 99.2  \\
               & B & Y & 89.9  & 98.1  & 99.2  \\
+ k=5                         & B & Y & 90.4  & --    & --    \\
\hline
\end{tabular}
\caption{Results on Mewsli-9 compared to the models described by~\citep{botha2020entity} and~\citep{decao2021multilingual}. Column~I indicates what is being indexed (Descriptions, Mentions, Both), and the HN indicates if additional rounds of Hard Negative training are applied.}
\label{tab:mewsli-overview}
\end{table}

\subsubsection{Per-Language Results}

Table~\ref{tab:mewsli-language} shows per-language results for Mewsli-9.
A key motivation of~\newcite{botha2020entity} was to learn a massively multilingual entity linking system, with a shared context encoder and entity representations between 104 languages in the Wikipedia corpus.
\moleman{} takes a step further: the indexed mentions from all languages are included in the retrieval index, and can contribute to the prediction in any language.
In fact, we find that for 21.4\% of mentions in the Mewsli-9 corpus, \moleman{}'s top prediction came from a different language.

\begin{table}[]
\centering
\small
\begin{tabular}{r|ccc}
\hline
Language  & R@1  & R@10 & R@100 \\
\hline
ar        & +1.1 & +0.9 & +0.3  \\
de        & -0.1 & +1.5 & +0.5  \\
en        & +0.3 & +2.8 & +2.3  \\
es        & -0.2 & +1.1 & +0.4  \\
fa        & +1.1 & +0.9 & +0.9  \\
ja        & +0.8 & +1.2 & +0.5  \\
sr        & -0.1 & +0.8 & +0.5  \\
ta        & +3.7 & +1.3 & +0.6  \\
tr        & +0.6 & +1.2 & +0.4  \\
\hline
micro-avg & +0.2 & +1.6 & +1.0  \\
macro-avg & +0.8 & +1.3 & +0.7 
\end{tabular}
\caption{\moleman{} results on the Mewsli-9 dataset by language, listed as a delta against \modelfplus{}~\citep{botha2020entity}.}
\label{tab:mewsli-language}
\end{table}

\subsubsection{Frequency Breakdown}

Table~\ref{tab:mewsli-frequency} shows a breakdown in performance by entity frequency bucket, defined as the number of times an entity was mentioned in the Wikipedia training set.
When indexing only mentions, \moleman{} can never predict the entities in the 0 bucket, but it shows significant improvement in the other frequency bands, particularly in the ``few shot'' bucket of [1,10).
This suggests when introducing new entities to the index, labelling a small number of mentions may be more beneficial than producing a single description.
To further confirm this intuition, we retrained \moleman{} with a modified training set which had all entities in the [1, 10) band of Mewsli-9 removed, and only added to the index at inference time.
This model achieved +0.2 R@1 and +5.6 R@10 relative to~\modelfplus{} (which was trained with these entities in the train set).
When entity descriptions are added to the index, \moleman{} outperforms \modelfplus{} across frequency bands.

\begin{table*}[]
\centering
\begin{tabular}{l|cc|cc||c|}
            & \multicolumn{2}{c|}{\textbf{\begin{tabular}[c]{@{}c@{}}MOLEMAN\\ (mentions only)\end{tabular}}} & \multicolumn{2}{c|}{\textbf{\begin{tabular}[c]{@{}c@{}}MOLEMAN\\ (+ descriptions)\end{tabular}}}  & \textbf{mGENRE} \\
            \hline
Freq. bin   & R@1           & R@10           & R@1  & R@10  & R@1                                          \\
            \hline
{[}0, 1)    & -8.3$\dagger$ & -33.9$\dagger$ & -0.2 & +18.3 & +13.8                                         \\
{[}1, 10)   & +0.4          & +5.6           & +1.7 & +9.3  & -10.4                                        \\
{[}10, 100) & +1.9          & +3.8           & +1.7 & +3.7  & -3.1                                        \\
{[}100, 1k) & +0.1          & +1.8           & -0.0 & +1.9  & +0.3                                        \\
{[}1k, 10k) & -1.1          & +0.7           & -1.2 & +0.7  & +0.6                                        \\
{[}10k,+)   & +0.7          & +0.6           & +0.7 & +0.5  & +2.2                                        \\
\hline
macro-avg   & -1.1          & -3.6           & +0.5 & +5.7  & +0.6                                        
\end{tabular}
\caption{Results from \moleman{} (with and without the inclusion of entity descriptions) on the Mewsli-9 dataset, by entity frequency in the training set plotted as a delta against \modelfplus{}. $\dagger$Note that when using mentions only, \moleman{} scores zero on entities that do not appear in the training set.}
\label{tab:mewsli-frequency}
\end{table*}

\subsubsection{Inference Efficiency}

Due to the large size of the mention index, nearest neighbor inference is performed using distributed maximum inner-product search.
We also experiment with approximate search using ScaNN~\citep{avq_2020}.
Table~\ref{tab:profile} shows throughput and recall statistics for brute force search as well as two approximate search approaches that run on a single multi-threaded CPU, showing that inference over such a large index can be made extremely efficient with minimal loss in recall.

\begin{table}[]
    \footnotesize
    \centering
    \begin{tabular}{c|c|c|c|c}
                     & QPS & Latency (ms) & R@1 & R@100 \\  
         \hline
         Brute-force & ~9.5 & 5727 & 89.9 & 99.2 \\
         ScaNN & 8000 & 2.9 & 89.9 & 99.1 \\
    \end{tabular}
    \caption{Max throughput (queries per second), latency (ms per query) and recall for brute force inference and approximate MIPS inference using the ScaNN library~\citep{avq_2020}. See Appendix~\ref{sec:profiling-details} for further details.}
    \label{tab:profile}
\end{table}

\subsection{Tsai Roth 2016 Hard}

In order to compare against previous multilingual entity linking models, we report results on the ``hard'' subset of~\newcite{tsai-roth-2016-cross}'s cross-lingual dataset which links 12 languages to English Wikipedia.
Table~\ref{tab:trhard} shows our results on the same 4 languages reported by~\newcite{botha2020entity}.
\moleman{} outperforms all previous systems.

\begin{table}[]
\centering
\begin{tabular}{r|ll}
    & MF+  & MM   \\
    \hline
de  & 0.62 & 0.64 \\
es  & 0.58 & 0.59 \\
fr  & 0.54 & 0.58 \\
it  & 0.56 & 0.59 \\
\hline
Avg & 0.57 & 0.60
\end{tabular}
\caption{Accuracy results on the TR2016$^{hard}$ test set for \modelfplus{} (MF+) and \moleman{} (MM)}
\label{tab:trhard}
\end{table}

\subsection{TACKBP 2010}

Recent work on entity linking have employed dual-encoders primarily as a retrieval step before reranking with a more expensive cross-encoder~\citep{wu2019scalable,agarwal2020entity}.
Table~\ref{tab:tackbp} shows results on the extensively studied TACKBP 2010 dataset~\citep{ji2010overview}.
\newcite{wu2019scalable} used a 24-layer BERT-based dual-encoder which scores the 5.9 million entity descriptions from English Wikipedia, followed by a 24-layer cross-encoder reranker.
\moleman{} does not achieve the same level of top-1 accuracy as their full model, as it lacks the expensive cross-encoder reranking step, but despite using a single, much smaller Transformer and indexing the larger set of entities from multilingual Wikipedia, it outperforms this prior work in retrieval recall at 100.

We also report the accuracy of a MOLEMAN model trained only with English training data, and using an Enlish-only index for inference.
This experiment shows that although the multilingual index contributes to~\moleman{}'s overall performance, the pairwise training data is sufficient for high performance in a monolingual setting.

\begin{table}[]
\centering
\begin{tabular}{lll}
\textbf{Method} & \textbf{R@1} & \textbf{R@100} \\
\hline
AT-Prior                        & -- & 89.5                \\
AT-Ext                          & -- & 91.7                \\
BM25                            & -- & 68.9                \\
\newcite{gillick2019learning}        & -- & 96.3                \\
\newcite{wu2019scalable}        & 91.5$\dagger$ & 98.3$^\ast$                \\
\hline
MOLEMAN (EN-only)               & 85.8 & 98.4 \\
MOLEMAN                         & 87.9 & 99.1               
\end{tabular}
\caption{Retrieval comparison on TACKBP-2010. The alias table and BM25 baselines are taken from~\newcite{gillick2019learning}.
For comparison to \newcite{wu2019scalable}, we report R@1 for their ``full Wiki, w/o finetune'' cross-encoder.
Their R@100 model is a dual-encoder finetuned on the TACKBP-2010 training set. \moleman{} is not finetuned.
}
\label{tab:tackbp}
\end{table}

\section{Discussion and Future Work}

We have recast the entity linking problem as an application of a more generic mention encoding task.
This approach is related to methods which perform clustering on test mentions in order to improve inference~\citep{le2018improving,angell2020clustering}, and can also be viewed as a form of cross-document coreference resolution~\citep{rao2010streaming,shrimpton-etal-2015-sampling,barhom2019revisiting}.
We also take inspiration from recent instance-based language modelling approaches~\citep{khandelwal2019generalization,lewis2020retrieval}.

Our experiments demonstrate that taking an instance-based approach to entity-linking leads to better retrieval performance, particularly on rare entities, for which adding a small number of mentions leads to superior performance than a single description.
For future work, we would like to explore the application of this instance-based approach to entity knowledge related tasks~\citep{seo2018phrase,petroni2020kilt}, and to entity discovery~\citep{ji2017overview}.

\section*{Acknowledgements}

The authors would like to thank Ming-Wei Chang, Livio Baldini-Soares and the anonymous reviewers for their helpful feedback.
We also thank Dave Dopson for his extensive help with profiling the brute-force and approximate search inference.

\bibliography{main}
\bibliographystyle{acl_natbib}

%

%
%
\appendix
\section{Appendices}
\label{sec:supplemental}

\subsection{Training setup and hyperparameters}

To isolate the impact of representing entities with multiple mention embeddings, we follow the training methodology and hyperparameter choices presented in~\newcite{botha2020entity} (Appendix A).

We train MOLEMAN using in-batch sampled softmax~\citep{gillick2018end} using a batch size of 8192 for 500k steps, which takes about a day.
Our model is implemented in Tensorflow~\citep{abadi2016tensorflow}, using the Adam optimizer~\citep{kingma2014adam,loshchilov2017decoupled} with the mention encoder preinitialized from a multilingual BERT checkpoint\footnote{\url{github.com/google-research/bert/multi_cased_L-12_H-768_A-12}}. All model training was carried out on a Google TPU v3 architecture\footnote{\url{cloud.google.com/tpu/docs/tpus}}.

\subsection{Datasets Links}

\begin{itemize}
    \item Mewsli-9: \url{http://goo.gle/mewsli-dataset}
    \item TR2016$^hard$: \url{cogcomp.seas.upenn.edu/
page/resource_view/102}
    \item TACKBP-2010: \url{https://catalog.ldc.upenn.edu/LDC2018T16}
\end{itemize}

\subsection{Profiling Details}
\label{sec:profiling-details}

The brute-force numbers we’ve reported are the theoretical maximum throughput for computing 300D dot-products on an AVX-512 processor running at 2.2Ghz, and are thus an overly optimistic baseline. Practical implementations, such as the one in ScaNN, must also compute the top-k and rarely exceed 70\% to 80\% of this theoretical limit. The brute-force latency figure is the minimum time to stream the database from RAM using 144 GiB/s of memory-bandwidth. In practice, we ran distributed brute-force inference on a large cluster of CPUs, which took about 5 hours.

The numbers for ScaNN are empirical single-machine benchmarks of an internal solution that uses the open-source ScaNN library~\footnote{\url{https://github.com/google-research/google-research/tree/master/scann}} on a single 24-core CPU. We use ScaNN to search a multi-level tree that has the following shape: $78,000 => 83:1 => 105:1$ (687.3 million datapoints). We used a combination of several different anisotropic vector quantizations that combine 3, 6, 12, or 24 dimensions per 4-bit code, as well as re-scoring with an \texttt{int8}-quantization.

\subsection{Expanded experimental results}

Tables~\ref{tab:mewsli-language-full} and ~\ref{tab:mewsli-frequency-full} present complete numerical comparisons between \moleman{} and \modelfplus{} on Mewsli-9.

\begin{table*}[h]
\centering
\begin{tabular}{r|ccc|ccc|ccc}
\multicolumn{1}{c}{}                  & \multicolumn{3}{c}{\textbf{Model F+}} & \multicolumn{3}{c}{\textbf{MOLEMAN}} & \multicolumn{3}{c}{\textbf{MOLEMAN}} \\ 
\multicolumn{1}{c}{}                  & \multicolumn{3}{c}{}                  & \multicolumn{3}{c}{\textbf{(mentions only)}} & \multicolumn{3}{c}{\textbf{(+ descriptions)}} \\ 
\hline
\multicolumn{1}{c}{\textbf{Language}} & R@1        & R@10       & R@100       & R@1        & R@10       & R@100     & R@1        & R@10       & R@100      \\
\hline
ar  & 92.3        & 97.7       & 99.1        & 93.4                      & 98.6                       & 99.0                         & 93.4                       & 98.6                       & 99.4                       \\
de  & 91.5        & 97.3       & 99.0        & 91.3                      & 98.2                       & 98.9                         & 91.5                       & 98.9                       & 99.5                       \\
en  & 87.2        & 94.2       & 96.7        & 87.4                      & 95.9                       & 97.4                         & 87.4                       & 97.0                       & 99.3                       \\
es  & 89.0        & 97.4       & 98.9        & 88.7                      & 98.1                       & 98.8                         & 88.7                       & 98.5                       & 99.3                       \\
fa  & 91.8        & 97.4       & 98.7        & 93.5                      & 98.5                       & 99.1                         & 92.9                       & 98.3                       & 99.6                       \\
ja  & 87.8        & 95.6       & 97.6        & 88.7                      & 96.2                       & 97.0                         & 88.5                       & 96.8                       & 98.0                       \\
sr          & 92.6        & 98.2       & 99.2        & 92.2                      & 98.7                       & 99.5                         & 92.5                       & 99.0                       & 99.7                       \\
ta          & 87.6        & 97.4       & 98.9        & 91.5                      & 98.4                       & 99.1                         & 91.3                       & 98.6                       & 99.5                       \\
tr          & 88.9        & 95.4       & 97.3        & 88.7                      & 97.6                       & 98.2                         & 89.1                       & 98.1                       & 99.2                       \\
\hline
micro-avg                      & 89.4        & 96.4       & 98.2        & 89.5                      & 97.4                       & 98.3                         & 89.6                       & 98.0                       & 99.2                       \\
macro-avg                      & 89.8        & 96.9       & 98.5        & 90.6                      & 97.8                       & 98.5                         & 90.6                       & 98.2                       & 99.3                      
\end{tabular}
\caption{Results on the Mewsli-9 dataset by language.}
\label{tab:mewsli-language-full}
\end{table*}

\begin{table*}[h]
\centering
\begin{tabular}{l|r|ccc|ccc|ccc}
\multicolumn{2}{c}{} & \multicolumn{3}{c}{Model F+} & \multicolumn{3}{c}{MOLEMAN} & \multicolumn{3}{c}{MOLEMAN} \\
\multicolumn{2}{c}{} & \multicolumn{3}{c}{} & \multicolumn{3}{c}{(mentions only)} & \multicolumn{3}{c}{(+description)} \\
            \hline
Bin         & Queries              & R@1     & R@10    & R@100    & R@1     & R@10    & R@100  & R@1     & R@10    & R@100 \\
{[}0, 1)    & 3,198                         & 8.3     & 33.9    & 62.7     & 0.0                  & 0.0                  & 0.0                  & 8.1     & 52.2    & 74.7    \\
{[}1, 10)   & 6,564                         & 57.7    & 80.8    & 91.3     & 58.1                 & 86.4                 & 93.3                 & 59.4    & 90.1    & 96.5    \\
{[}10, 100) & 32,371                        & 80.4    & 92.8    & 96.7     & 82.2                 & 96.5                 & 98.8                 & 82.1    & 96.5    & 98.9    \\
{[}100, 1k) & 66,232                        & 89.6    & 96.6    & 98.2     & 89.7                 & 98.4                 & 99.5                 & 89.6    & 98.5    & 99.5    \\
{[}1k, 10k) & 78,519                        & 92.9    & 98.4    & 99.3     & 91.9                 & 99.2                 & 99.8                 & 91.8    & 99.1    & 99.8    \\
{[}10k, +)  & 102,203                       & 94.1    & 98.8    & 99.4     & 94.8                 & 99.4                 & 99.6                 & 94.8    & 99.3    & 99.5    \\
\hline
micro-avg   & \multicolumn{1}{l}{}          & 89.4    & 96.4    & 98.2     & 89.5                 & 97.4                 & 98.3                 & 89.6    & 98.0    & 99.2    \\
macro-avg   & \multicolumn{1}{l}{}          & 70.5    & 83.5    & 91.3     & 69.4                 & 80.0                 & 81.8                 & 70.9    & 89.3    & 94.8 
\end{tabular}
\caption{Results on the Mewsli-9 dataset, by entity frequency in the test set.}
\label{tab:mewsli-frequency-full}
\end{table*}
%
%


\end{document}